\documentclass[12pt,preprint]{aastex}
\def\gsim{{{}_>\atop{}^{{}^\sim}}} 
\def\lsim{{{}_<\atop{}^{{}^\sim}}}
\def\eq#1{equation~(\ref{#1})}

\def\keywords{}

\begin{document}

\def\newpage{\vfill\eject} 
\def\vs{\vskip 0.2truein}
\def\msun{M_\odot} 
\def\rsun{R_\odot} 
\def\met{[M/H]} 
\def\vi{(V-I)}
\def\mtot{M_{\rm tot}} 
\def\mhalo{M_{\rm halo}} 
\def\pp{\parshape 2 0.0truecm 16.25truecm 2truecm 14.25truecm}
\def\la{\mathrel{\mathpalette\fun <}}
\def\ga{\mathrel{\mathpalette\fun >}}
\def\fun#1#2{\lower3.6pt\vbox{\baselineskip0pt\lineskip.9pt
\ialign{$\mathsurround=0pt#1\hfil##\hfil$\crcr#2\crcr\sim\crcr}}}
\def\kpc{{\rm kpc}} 
\def\Mpc{{\rm Mpc}} 
\def\kmsec{{\rm km/sec}}
\def\ibl{{\cal I}(b,l)} 
\def\kms{{\rm km}\,{\rm s}^{-1}}
\def\dii{\delta I/I} 
\def\di0{(\delta I/I)_0} 
\def\tp{t_{\rm peak}}
\def\days{\rm days} 
\def\tp{t_{\rm peak}}
\def\td{{t_{\rm day}}} 
\def\thetae{\theta_{\rm E}} 
\def\ftot{F_{\rm tot}} 
\def\nr{N_{\rm r}} 
\def\ci{{\cal I}} 
\def\taus{\tau_{\rm start}}
\def\taue{\tau_{\rm end}}
\def\tb{t_{\rm b}}
\def\num{\nu_{\rm m}}

\title{Microlensing and the Surface Brightness Profile \\
of the Afterglow Image of GRB 000301C}

\author{B. Scott Gaudi\footnote{Hubble Fellow}} 
\affil{Institute for Advanced Study, Einstein Drive, Princeton, NJ 08540}
\authoremail{gaudi@sns.ias.edu} 

\author{Jonathan Granot} 
\affil{Racah Institute of Physics, Hebrew University, Jerusalem, 91904,
Israel} 
\authoremail{jgranot@nikki.fiz.huji.ac.il}

\author{Abraham Loeb}
\affil{Harvard-Smithsonian CfA, 60 Garden Street, Cambridge, MA 02138}
\authoremail{aloeb@cfa.harvard.edu}



\begin{abstract}
The optical afterglow of Gamma-Ray Burst (GRB) 000301C exhibited a
significant, short-timescale deviation from the power-law flux
decline expected in the standard synchrotron shock model.
Garnavich, Loeb \& Stanek found that this deviation was well-fit by
an {\it ad hoc} model in which a thin ring of emission is
microlensed by an intervening star.  We revisit the microlensing
interpretation of this variability, first by testing whether
microlensing of afterglow images with realistic surface brightness
profiles (SBPs) can fit the data, and second by directly inverting
the observed light curve to obtain a non-parametric measurement of
the SBP.  We find that microlensing of realistic SBPs can reproduce
the observed deviation, provided that the optical emission arises
from frequencies above the cooling break.
Conversely, if the variability is indeed caused by microlensing, the SBP
must be significantly limb-brightened. Specifically, $\ge 60\%$ of the flux
must originate from the outer $25\%$ of the area of the afterglow image.
The latter requirement is satisfied by the best fit theoretical SBP.
The underlying optical/infrared afterglow lightcurve is consistent
with a model in which a jet is propagating into a uniform medium with
the cooling break frequency below the optical band.

\end{abstract}


\setcounter{footnote}{0}
\renewcommand{\thefootnote}{\arabic{footnote}}

\section{Introduction\label{sec:intro}}

The afterglows of gamma-ray bursts (GRBs) are observed in the X-ray,
optical, near-infrared, and radio, and appear to be well-described by
the synchrotron blast-wave model in which the source ejects material
with a relativistic bulk Lorentz factor, driving a relativistic shock
into the external medium (see \citealt{vp2000,piran2000}
and references therein).  There is mounting evidence from
the observed steepening of afterglow light curves that these ejecta
are in many cases mildly to highly collimated, with opening angles
$\sim 3$--$30^\circ$ \citep{harrison1999,fwk2000,frail2001}.  Global
fitting of the afterglow light curves over many decades in time and
frequency, in the context of this model, can be used to derive
constraints on the physical parameters of the model, i.e., the energy
and opening angle of the jet, the external density, the magnetic field
strength and the energy distribution of the electrons behind the shock
\citep{wg1999,fandw1999,pandk2001}.

In this model, the image of the afterglow is expected to appear highly limb
brightened at frequencies above the peak synchrotron frequency $\num$, but
more uniform at frequencies $< \num$, especially below the self--absorption
frequency $\nu_a$ (Waxman 1997; Sari 1998; Panaitescu \& M\'esz\'aros 1998;
Granot, Piran, \& Sari 1999a,b; Granot \& Loeb 2001).
A measurement of the surface brightness profile (SBP) at several frequencies
would thus provide an important test of the model.  For typical parameters,
the afterglow image expands superluminally, and has an angular radius
${\cal O}(\mu{\rm as})$ a few days after the GRB.  As pointed out by
\citet{landp1998}, this is of the same order as the angular Einstein ring
radius of a solar mass lens at cosmological distances,
\begin{equation}
\thetae=\left( {4GM \over c^2D}\right)^{1/2} = 1.6 \left({M \over
M_\odot}\right)^{1/2}\left({D \over 10^{28}~{\rm cm}}\right)^{-1/2}
\mu{\rm as},
\label{eqn:thetae}
\end{equation}
where $M$ is the lens mass, and $D\equiv D_{os}D_{ol}/D_{ls}$, and
$D_{os}$, $D_{ol}$ and $D_{ls}$ are the angular diameter distances between
the observer-source, observer-lens, and lens-source, respectively.  Thus
lensing by a star along the line of sight will produce a detectable
magnification over the course of a few days to weeks.  The probability that
any given GRB will be microlensed is $\sim 1\% b^2$, where $b$ is the
angular separation between the GRB and lens in units of $\thetae$.  Since
the original suggestion by \citet{landp1998}, this application of
microlensing has been studied by several authors
\citep{mandl2001,kandw2001,gl2001,gandl2001,iandn2001}.

The well-sampled optical afterglow of GRB 000301C exhibited a short time
scale, nearly achromatic deviation from the nominal power-law flux decline.
Garnavich, Loeb \& Stanek (2000; hereafter GLS) found that this 
deviation is well-fit by 
a model in which the afterglow was microlensed by an
intervening star.  However, the model they adopted for the SBP was somewhat
crude: they assumed the emission arose solely from an outer ring of
fractional width $W$, with $W\simeq 10\%$ providing the best fit to the data.
Recently, \citet{pana2001} argued that, if one adopts
realistic SBPs, microlensing cannot explain the data.  Here we revisit
the microlensing interpretation of this deviation.  We fit the
optical/infrared dataset of GRB 000301C to a double-power law plus
microlensing model, adopting realistic SBPs, and, under the assumption
that the observed deviation is indeed due to microlensing, we directly
invert the light curve, obtaining a non-parametric
measurement of the SBP, and compare this measurement to theoretical
expectations.

\begin{figure*}   
\epsscale{1.0} \centerline{\plotone{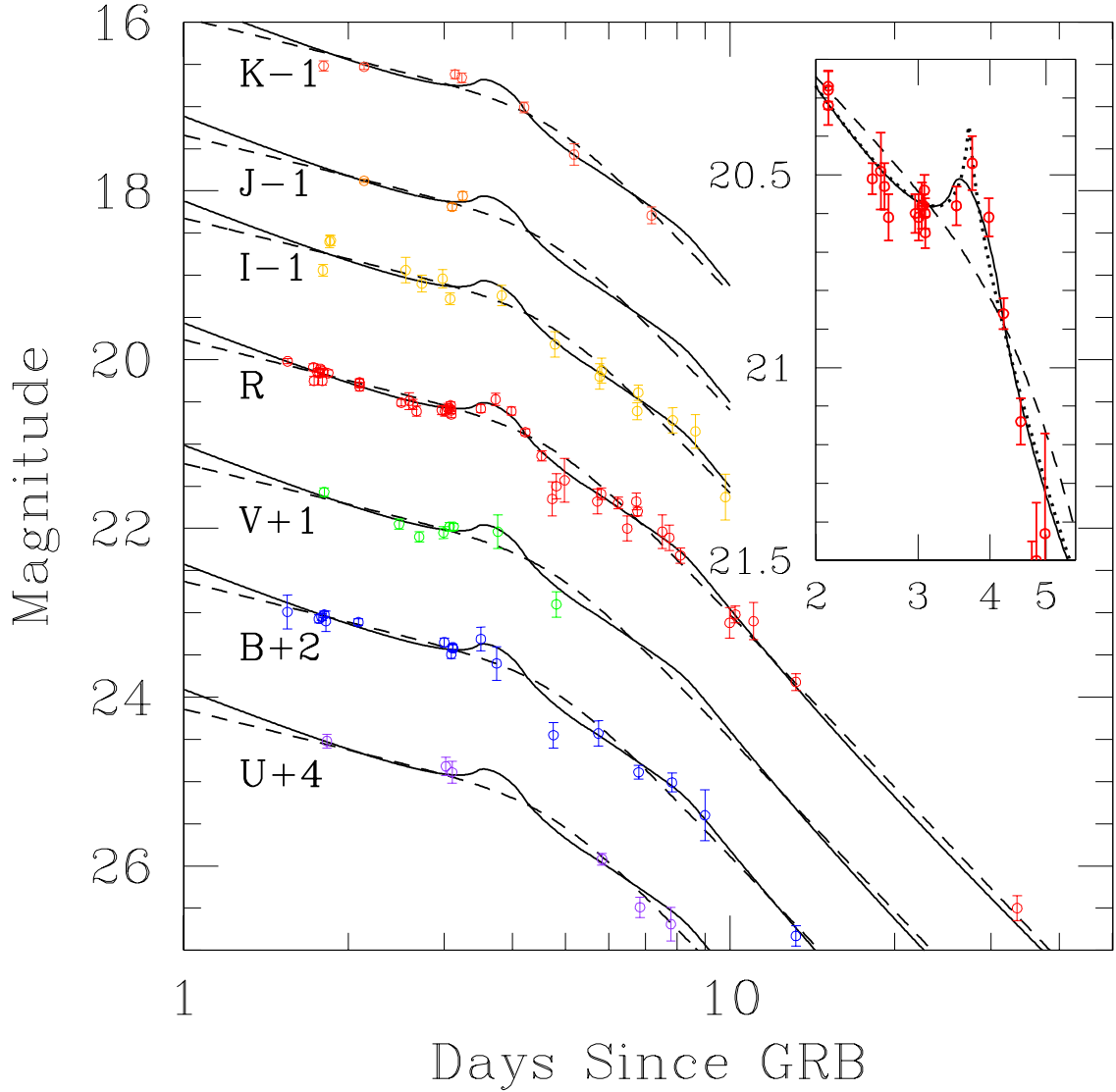}}
\caption{
$UBVRIJK$ photometry of GRB 000301C as a function of days from the
GRB.  Points have been offset by the indicated amount for clarity.
The dashed line is the best-fit double power law (with no lensing),
while the solid line is the overall best-fit microlensing model, where
the SBP has been determined from direct inversion.  The inset shows
the $R$-band data only.  The dotted line is the best-fit microlensing
model with theoretically calculated SBP, in this case for $k=0$ and
$\nu>\nu_c$.  }
\label{fig:fig1}
\end{figure*}

\section{Data}

We will be considering the same dataset as GLS, namely 104 photometric
data points in the $UBVRIJK$-bands distributed as $(6, 18, 8, 46, 16,
3, 7)$.  These data, shown in Figure 1, are taken from the compilation
of \citet{sagar2000} with additional photometry from
\citet{stanek2000}.  We will not consider the data in the radio regime
\citep{berger2000}, as the majority is not contemporaneous with the
optical deviation.  For more details on the GRB itself and the
optical, infrared and radio data, see GLS, \citet{masetti2000}, \citet{sagar2000},
\citet{berger2000} and \citet{randf2000}.

\section{Theoretical Surface Brightness Profiles}\label{SBPs}

The magnitude, shape, and duration of the microlensing signal depends
sensitively on the SBP of the afterglow
\citep{landp1998,mandl2001,gl2001}.  The SBP, in turn, depends on the
observed frequency and the physical parameters of the shock.  Thus
detailed measurements of a microlensed afterglow light curve can be
used to constrain the physical parameters of the afterglow. However,
the expected SBPs in the relativistic blast-wave model can also be
calculated {\it a priori} under various assumptions
\citep{wax1997c,sari1998,gps1999,gps1999b,gl2001,iandn2001,pana2001}.
Thus one can alternatively use the theoretical SBPs to calculate the
expected microlensing signature, thereby confirming or refuting the
microlensing interpretation\footnote{Since both the unlensed light
curves and the magnification history depend on the physical assumptions
in the afterglow model, a poor fit for a certain microlensing model
might suggest an unrealistic afterglow model, rather than refute the
microlensing interpretation altogether.}.
This was the approach taken by \citet{pana2001}.  We adopt both
approaches here.

\begin{figure*}   
\epsscale{1.0} \centerline{\plotone{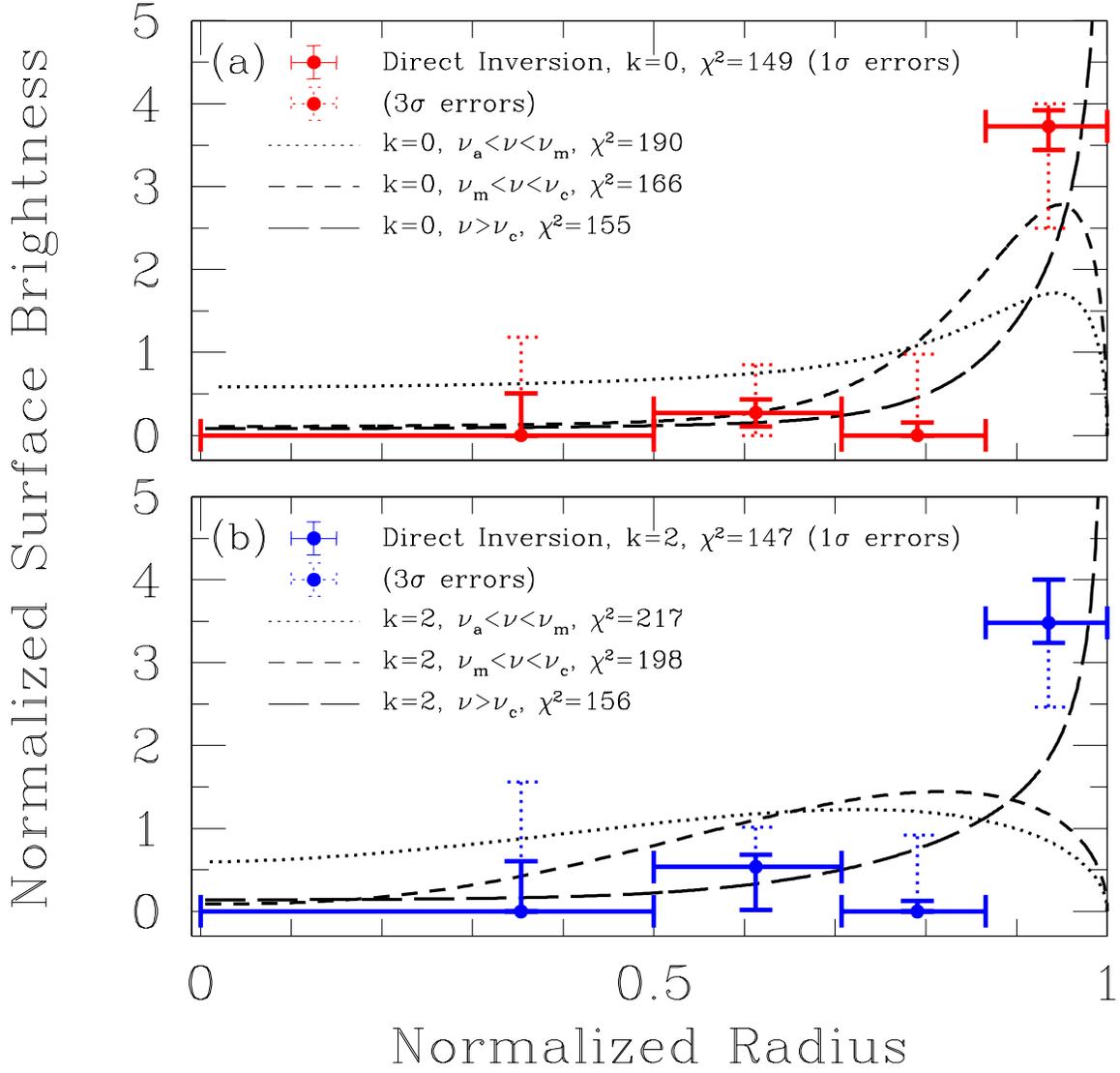}}
\caption{SBPs as a function of normalized radius.  The points are the SBPs
determined from direct inversion, with $1\sigma$ errors (solid) and
$3\sigma$ errors (dotted).  The curves are theoretically calculated SBPs
for various frequency regimes (see Granot \& Loeb 2001).  (a) Uniform
external medium, $k=0$.
(b) Stellar wind environment, $k=2$.}
\end{figure*}

We calculate SBPs using the method described in \citet{gs2001} and Granot
\& Loeb (2001).  Briefly, the hydrodynamics are described by the
Blandford-McKee (1976; BM hereafter) self-similar solution, assuming a
power-law external density profile, $\rho(r) \propto r^{-k}$, and a
power-law number versus energy distribution of electrons with index
$p$, ${\rm d}N_e/{\rm d}\gamma\propto \gamma^{-p}$, just behind the
shock, which thereafter evolves due to radiative and adiabatic losses.
We integrate the emission over the entire volume behind the shock
front.  A few days after the burst the afterglow is typically in the
slow cooling regime (i.e.\ a typical electron cools on a timescale
larger than the dynamical time) and the ordering of the break
frequencies is $\nu_a<\nu_m<\nu_c$, where $\nu_c$ is the cooling
frequency \citep{spn1998,gps1999,cl2000}.  We assume that this is the
case for GRB 000301C, as indicated by \citet{berger2000}.
We consider only the SBPs from frequencies $\nu > \nu_a$, since after
a few days the optical band is typically well above $\nu_a$ (and
usually also above $\nu_m$). The spectrum consists of several
power-law segments (PLSs), where $F_{\nu}\propto\nu^{\beta}t^{\alpha}$
(see Figure 1 in Granot \& Loeb 2001). Due to the self-similar nature
of the hydrodynamics, the normalized SBP within a given PLS is
independent of time (though it changes considerably between different
PLSs). Thus, for each $k$, there are 3 relevant forms for the SBP,
each corresponding to a different PLS: $\beta=1/3$ ($\nu_a < \nu <
\nu_m$), $\beta=(1-p)/2$ ($\nu_m < \nu <
\nu_c$) and $\beta=-p/2$ ($\nu > \nu_c$). The corresponding values of the
temporal index $\alpha$ for $k=0$ ($k=2$) are $\alpha=1/2$ ($\alpha=0$) for
$\beta=1/3$, $\alpha=-3(p-1)/4$ ($\alpha=-(3p-1)/4$) for $\beta=(1-p)/2$
and $\alpha=-(3p-2)/4$ (for both $k=0$ and $k=2$) for $\beta=-p/2$.

The expected SBPs are shown in Figure 2, for a uniform external medium
$k=0$, and a stellar wind environment $k=2$. Note that the SBP becomes more
limb-brightened as the frequency goes to a higher PLS, i.e.\ the spectral
slope $\beta$ is reduced, or as the value of $k$ is reduced.

\section{Analysis and Model Fitting}

We fit the data of the flux as a function of time to the form of two
asymptotic power-laws which join smoothly:
\begin{equation}
F_\nu(t)=\mu(t) F_{0,\nu} 
\left[ \left({t\over\tb}\right)^{-s\alpha_1}+
\left({t\over\tb}\right)^{-s\alpha_2}\right]^{-1/s},
\label{eqn:foft}
\end{equation}
where $\tb$ is the ``break time,'' $F_{0,\nu}$ is the unlensed flux at
the point where the asymptotic power laws before and after the break
meet, $\alpha_1$ and $\alpha_2$ are the indices of the flux decline,
and $s$ is a sharpness parameter; the larger the value of $s$, the
sharper the break. The magnification $\mu$ due to microlensing depends
on the impact parameter $b$,
the radial SBP of the afterglow image, and the angular size of the
afterglow in units of $\thetae$,
\begin{equation}
R_{\rm s}={R_0 \td^{\delta}}.~\label{eqn:rhot}
\end{equation}
Here $\delta=(5-k)/2(4-k)$, $\td$ is the time from the GRB in days and
$R_0$ is the value of $R_s$ after 1 day.  The light curve exhibits an
achromatic break at $\tb\sim 7$~days.  Berger et al.\ (2000) find that
this break can be well explained by jet with $k=0$, while Panaitescu
(2001) argues that a jet alone cannot produce the required steepening
of the light curve, and invokes a double power law electron
distribution, which causes additional (though chromatic) steepening of
the light curve if the synchrotron frequency corresponding to the
break in the electron distribution crosses the optical band around
$\tb$.  Either way, we would expect the SBP at $t\gsim\tb$ to be
different than at $t\lsim\tb$. However, our best fit models give
$\tb\sim 8~$days, while the majority of the constraints on the SBP
come from times $t \lsim 8~$days, where the spherical self-similar BM
solution should still be applicable, resulting in the self-similar
SBPs described in \S \ref{SBPs}.  We will therefore adopt the simpler
(and self consistent) working assumption that the SBP, when normalized
to its average value, as a function of radius normalized to $R_{\rm
s}$, is independent of time within a given PLS.

The model in \eq{eqn:foft} is a function of $N_{\rm par}=4+N_l+2$ free
parameters: four parameters describe the shape of the broken power-law
($\alpha_1, \alpha_2, s, \tb$); one parameter provides the
normalization $F_{0,\nu}$ for each of $N_l$ frequency
bands\footnote{Standard afterglow models predict
$F_{0,\nu}\propto\nu^{\beta}$.  However, extinction from the Milky Way
and possibly the host galaxy itself will induce an (uncertain)
curvature in the spectrum.  Indeed, as noted by
\citet{randf2000,jensen2001} and \citet{pana2001}, the early optical/infrared
data for GRB 000301C show evidence
for a curved spectrum, which may be due to extinction or is perhaps
intrinsic.  For simplicity, we will use a free parameter for each
$F_{0,\nu}$.  We have also performed tests where we fit the data
assuming the form $F_{0,\nu}\propto\nu^{\beta}$.  In general we find
worse fits, though our conclusions regarding the SBPs are essentially
unchanged.  As we show, the results for the $R$-band data are the same
as for the full dataset, indicating that our assumption of independent
$F_{0,\nu}$ does not bias our results.}  and two additional parameters
describe the microlensing effect ($b,R_0$) for a given SBP.  Considering all the
data for GRB 000301C in Figure 1, there are $11$ parameters without
microlensing, and $13$ free parameters with microlensing using
theoretically-calculated SBPs.  
For comparison purposes, we also fit
the data to the model adopted by GLS: we assume that the source
emission is confined to a thin ring of fractional width $W$.  However,
we extend the model of GLS by allowing the image interior to the ring to
have a relative surface brightness $C$ referenced to the outer ring 
($C=0$ reduces to the GLS model).  This model introduces two
additional free parameters.  
Finally, we consider a non-parametric fit
(``direct inversion'') to the SBP. We divide the image into $N_{\rm
bin}$ equal-area annuli, and find the fraction $f_i$ of the total flux
contributed by each bin $i$.  In this case the magnification is given
by $\mu=\sum_i f_i \mu_i$, where $\mu_i$ is the magnification of
annulus $i$.  This adds an additional $N_{\rm bin}-1$ parameters (due
to the constraint that $\sum_i f_i = 1$).  We find the best-fit
solution by minimizing $\chi^2$ with respect to all of these
parameters.  We define the $1\sigma$ errors on these parameters as the
projection of the $\Delta \chi^2=1$ hypersurface on the parameter
axes, where
\begin{equation}
\Delta\chi^2 \equiv { {\chi^2 -\chi^2_{\rm min}} \over \chi^2_{\rm min}/{\rm dof}},
\label{eqn:dchi}
\end{equation}
$\chi^2_{\rm min}$ is the minimum $\chi^2$ for a given model, and
${\rm dof}=104-N_{\rm par}$ is the number of degrees of freedom
for the 104 data points.  We normalize $\Delta\chi^2$ by the factor
$\chi^2_{\rm min}/{\rm dof}$ because we believe that the errors on the data points 
are likely underestimated, resulting in inflated values of $\chi^2$.  
Errors on the fit parameters determined using these inflated values of $\chi^2$
would be significantly underestimated. 

\section{Results}

Figure 1 shows the best fit for the double power-law model with no
lensing.  The best-fit parameters and $1\sigma$ errors for $(\alpha_1,
\alpha_2, s, \tb)$ are tabulated in Table 1.  The fit is poor:
$\chi^2=240.7$ for $93$ dof. Residuals from the broken power-law model
are shown in Figure 3; the systematic deviations from this model are
apparent.

When microlensing is included, the fit improves considerably. 
For the extended GLS model, we find that the data is best explained by
emission solely from an outer ring of relatively small fractional
width.  Specifically, we find best-fit parameters,
$W=0.13^{+0.04}_{-0.07}$, $C=0^{+0.02}$ for $k=0$ and
$W=0.16_{-0.08}^{+0.03}$, $C=0^{+0.03}$ for $k=2$. The remaining
parameters and the best-fit $\chi^2$ are shown in Table 1.  The
parameters we find differ from those of GLS, and we find a
$\chi^2/{\rm dof}$ smaller than they report.  However, when we
evaluate their model fit, we recover their $\chi^2/{\rm dof}$,
indicating that their fit was likely a local, not global, minimum.
For the fits with microlensing, we find that the sharpness parameter
$s$ is not well constrained, but is always $s\gsim 5$,
indicating a sharp break.  To avoid excessive covariances with other
parameters, we fix $s=20$ for the remaining fits, thus reducing $N_{\rm
par}$ by one.  The results are essentially identical for other values of $s\gsim 5$.   For
a uniform external medium ($k=0$), for which one generally predicts more
limb-brightened SBPs, we find $\chi^2=190.0, 166.1$, and $154.7$ for
$\nu_a<\nu<\nu_m$, $\nu_m<\nu<\nu_c$, and $\nu>\nu_c$, respectively.  For
the case of a wind-like medium ($k=2$), the fits are generally worse:
$\chi^2=216.8, 197.9$, and $155.5$ for the same frequency ranges.  The
deviations of these fits from the double power-law model are shown in
Figure 3.  It is clear that a significantly limb-brightened profile is
required to fit the sharp, large deviation at $t\sim 4$~days.  For $\nu >
\nu_c$, both wind and uniform external media provide satisfactory fits to
the data.  For $\nu_m<\nu<\nu_c$, the uniform medium provides a marginal
fit.  Frequencies below the peak synchrotron frequency $\nu_m$, or
$\nu_m<\nu<\nu_c$ with a wind-like medium ($k=2$), result in SBPs that are
too uniform to fit the data.

For the non-parametric SBP fit to the data, we impose the condition
that $f_i \ge 0$, and we increase the number of bins $N_{\rm bin}$ in
the recovered SBP until $\chi^2$ changes by $\le 1$.  We find that
$N_{\rm bin}=4$ is optimal.  We find $\chi^2=148.5$ for $k=0$, and
$\chi^2=146.9$ for $k=2$, for 89 dof.  
The inferred surface brightnesses are shown in Figure 2, the fit for $k=0$
is shown in Figure 1, and Figure 3 shows the deviations of both the $k=0$
and $k=2$ models from the best-fit double power law.  The best-fit values
and $1\sigma$ errors for the parameters $\alpha_1$, $\alpha_2$, $\tb$,
$R_0$, and $b$ are given in Table 1 for all viable ($\Delta\chi^2 \le 20$)
fits.  Although the values of $\chi^2/{\rm dof}$ for the direct inversion
formally indicate poor fits, it is clear from Figures 1 and 3 that
microlensing reproduces the observed light curve shape quite faithfully, at
least for limb-brightened sources.  It is likely that underestimated errors
contribute significantly to the inflated $\chi^2$.  

We have also fit the $R$-band data only using both the SBP determined
from direct inversion and the theoretically-calculated SBPs.  The SBP
inferred from direct inversion is consistent with that found from
fitting all the data (with larger errors, of course). The fit is
better: $\chi^2/{\rm dof}=1.23$ for 37 dof.  Fits using the
theoretical SBP yield the same conclusions as the fits to the entire
data set: the SBPs for $\nu > \nu_c$ provide fits that are acceptable
($\chi^2/{\rm dof}=1.24$ and $1.26$ for $k=0$ and $k=2$ respectively,
for 40 dof).  The fit for $\nu_m<\nu < \nu_c$, and $k=0$ is marginal:
$\chi^2/{\rm dof}=1.68$.  All other fits are considerably worse:
$\chi^2/{\rm dof}>2$.
\begin{figure*}   
\epsscale{1.0} \centerline{\plotone{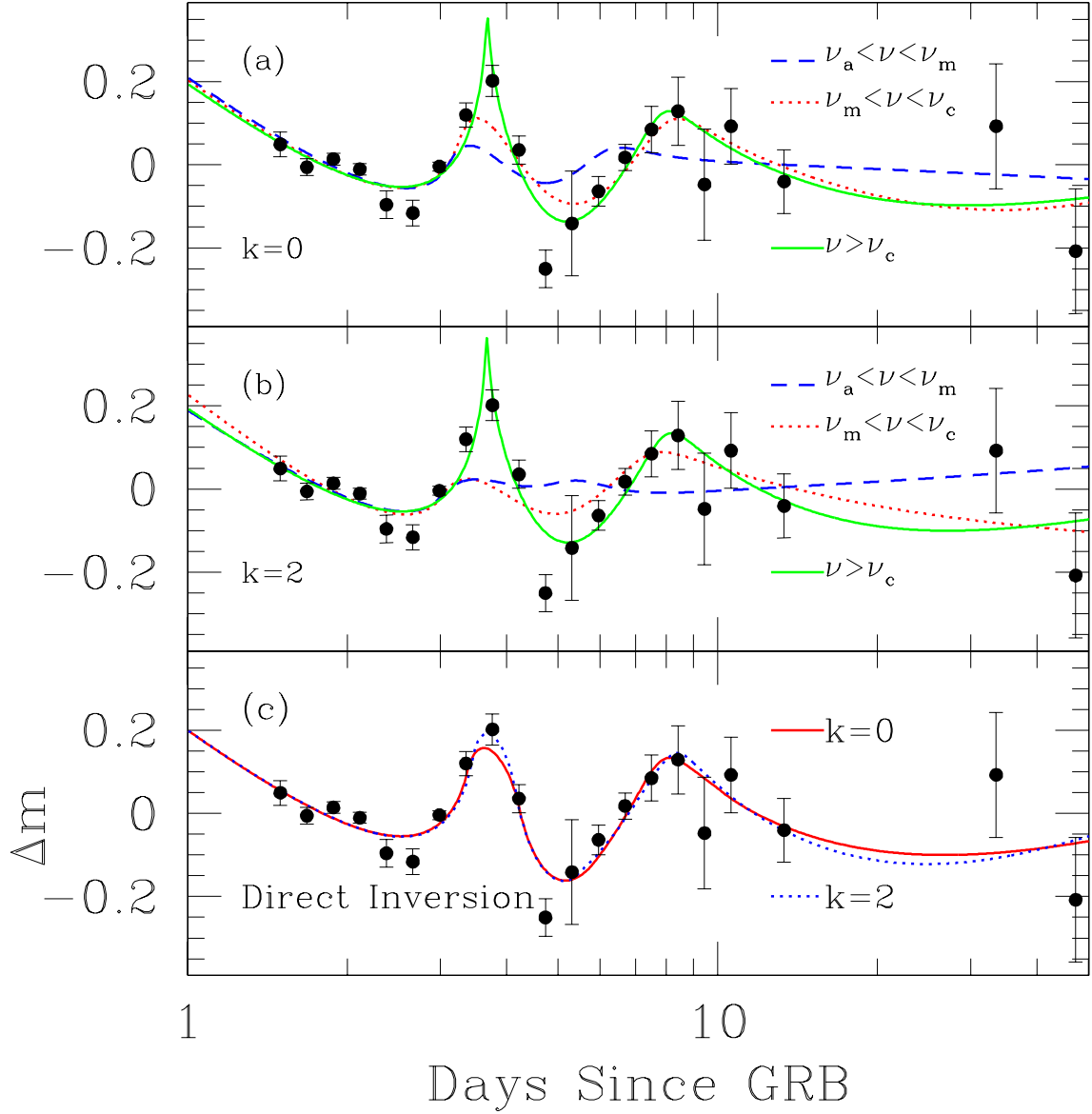}}
\caption{ In each panel, the points show the binned residuals from the
best-fit double power-law fit (without lensing) in magnitudes.  The curves
show the difference between microlensing models with various SBPs and the
double power law.  (a) Theoretically calculated SBPs for k=0. (b)
Theoretically calculated SBPs for k=2.
(c) SBPs determined from direct inversion.
}
\label{fig:fig3}
\end{figure*}

\begin{table*}[t]
\begin{center}
\begin{tabular}{|c||c|c|c|c|c|c|c|}
\tableline 
Model & $-\alpha_1$ & $-\alpha_2$ & $s$ & $\tb$ & $b$  & $R_0$ & $\chi^2$/dof \\ 
\tableline 
No $\mu$-lensing & $0.64_{-0.06}^{+0.05}$ & $2.72_{-0.10}^{+0.11}$ & $3.14_{-0.91}^{+1.37}$
& $4.80_{-0.19}^{+0.21}$ & --- & --- & 240.7/93 \\
$k=0$, $\nu_m<\nu<\nu_c$& $1.07_{-0.02}^{+0.02}$ & $2.31_{-0.10}^{+0.23}$ & 20\tablenotemark{1} & $8.14_{-0.48}^{+0.54}$ & $0.24_{-0.24}^{+0.23}$ & $0.12_{-0.12}^{+0.11}$ & 166.1/92\\ 
$k=0$, $\nu>\nu_c$& $1.06_{-0.02}^{+0.02}$ & $2.62_{-0.11}^{+0.10}$ & 20\tablenotemark{1} & $7.89_{-0.42}^{+0.49}$ & $0.65_{-0.13}^{+0.13}$ & $0.29_{-0.06}^{+0.06}$ & 154.7/92\\
$k=2$, $\nu>\nu_c$& $1.04_{-0.02}^{+0.02}$ & $2.63_{-0.11}^{+0.11}$ & 20\tablenotemark{1} & $8.01_{-0.44}^{+0.51}$ & $0.60_{-0.12}^{+0.12}$ & $0.23_{-0.05}^{-0.05}$ & 155.5/92\\
$k=0$, DI\tablenotemark{2} & $1.06_{-0.02}^{+0.02}$ & $2.60_{-0.10}^{+0.11}$ & 20\tablenotemark{1} & $7.88_{-0.41}^{+0.45}$ & $0.62_{-0.19}^{+0.24}$ & $0.29_{-0.06}^{+0.07}$ & 148.5/89\\
$k=2$, DI\tablenotemark{2} & $1.05_{-0.02}^{+0.02}$ & $2.51_{-0.10}^{+0.11}$ & 20\tablenotemark{1} & $8.36_{-0.48}^{+0.49}$ & $0.44_{-0.40}^{+0.41}$ & $0.18_{-0.18}^{+0.13}$ & 146.9/89\\
$k=0$, GLS\tablenotemark{3} & $1.06_{-0.02}^{+0.02}$ & $2.71_{-0.09}^{+0.10}$ & $15.5_{-10.0}^{+\infty}$ & $7.74_{-0.37}^{+0.37}$ & $0.87_{-0.14}^{+0.16}$ & $0.41_{-0.08}^{+0.07}$ & 147.3/89\\
$k=2$, GLS\tablenotemark{3}  & $1.05_{-0.02}^{+0.02}$ & $2.75_{-0.10}^{+0.10}$ & $13.8_{-7.8}^{+\infty}$ & $7.68_{-0.36}^{+0.41}$ & $0.92_{-0.12}^{+0.15}$ & $0.37_{-0.07}^{+0.06}$ & 146.9/89\\
\tableline
\end{tabular}
\end{center}
\tablenum{1} {\bf Table 1} Fit parameters and $1\sigma$ errors for GRB 000301C.
\tablenotetext{1} {Fixed.}
\tablenotetext{2} {Fit with the SBP determined from direction inversion.}
\tablenotetext{3} {Fit using a modified version of the SBP model adopted by GLS.}
\label{tbl:table1}
\end{table*}

\section{Discussion and Conclusions }

We find that the early-time ($t<\tb$) flux decline is well-constrained,
$\alpha_1=-1.05\pm 0.02$.  The slope of the optical spectrum at this time
is $\beta=-0.9$, taking into account Milky Way (MW) extinction only, but
can be as shallow as $\beta=-0.7$ if SMC-type host extinction is included
\citep{randf2000,jensen2001}.  For $\nu > \nu_c$ (the SBPs most consistent
with the data), the value of $\alpha_1$ implies $p\simeq 2.07$ and thus
$\beta=-1.03$, which is only slightly steeper than the observed spectrum
for MW extinction.  For $\nu_m < \nu < \nu_c$ and $k=0$, $\alpha_1=-1.05$
implies $p=2.4$ and thus $\beta=-0.7$, consistent with observations {\it
if} SMC-type host extinction is adopted.  If the break at $t\sim 8~{\rm
days}$ is due to a jet, then simple analytic models predict that the change
in temporal flux index $\alpha$ should be $\Delta\alpha=(p+3)/4$ for $k=0$
and $\nu_m < \nu < \nu_c$ or $\Delta\alpha=(p+2)/4$ for $\nu>\nu_c$ and
$k=0$ or $2$ \citep{sph1999,pandk2001}. Kumar \& Panaitescu (2000) find,
using a semi-analytic model, that the jet break is rather smooth,
especially for a wind environment ($k=2$), and therefore inconsistent with
the sharp break that we infer. However, initial results of numerical
calculations of the jet break \citep{granot2001}, based on a 2D
hydrodynamic simulation of the jet for $k=0$, indicate that the break can
be rather sharp (with $s\sim 4.5$ for an observer along the jet axis), and
that $\alpha_2$ is smaller by $\sim 0.35$ compared to the simple analytic
predictions mentioned above, making $\Delta\alpha$ larger by a similar
factor.  Nevertheless, the sharpness of the break still presents a serious
problem for the model where the jet propagates into a stellar wind
($k=2$). Also, the SBP is expected to become more uniform at
$t>\tb$ compares to $t<\tb$ \citep{iandn2001,pana2001}. Since we
assume that the form of the SBP relevant for $t<\tb$ holds all along,
we underestimate the value of $\alpha_2$, and overestimate the value
of $\Delta\alpha$ accordingly\footnote{This effect may be seen in
Table 1, where $\alpha_2$ is larger for $\nu_m<\nu<\nu_c$ (for which
the SBP is more homogeneous), compared to $\nu>\nu_c$, for $k=0$ or
$2$, (for which the SBP is more limb brightened).}.  This effect is
expected to be larger for $\nu>\nu_c$, where the SBP is more limb
brightened to begin with. The values we obtain for $\Delta\alpha$ may
therefore serve as upper limits (especially for $\nu>\nu_c$). For the
SBP fit for $k=0$ and $\nu_m < \nu < \nu_c$, we find
$\Delta\alpha\lesssim 1.2_{-0.1}^{+0.2}$ whereas based on the electron
index derived from $\alpha_1$, the prediction is $\Delta\alpha\approx
1.7$.  For $\nu > \nu_c$ and $k=0$, we find $\Delta\alpha\lesssim
1.56\pm 0.11$, which is reasonably consistent with the prediction 
$\Delta\alpha\approx 1.37$.

Thus, assuming the break in the light curve is due to jet effects, the only
scenario which provides a satisfactory explanation of the data is
$\nu>\nu_c$ and $k=0$. For $\nu < \nu_c$, the inferred SBP is only
marginally consistent with the expected profile, and the observed change in
the flux indices is smaller than the theoretical expectation.  For $\nu >
\nu_c$ and $k=2$, the observed break is too sharp compared to theoretical
predictions. We note that our best fit model parameters, $\nu>\nu_c$ and
$k=0$, are also favored by \citet{pana2001}, and said to provide a good fit
to the data by \citet{berger2000}.

\citet{pana2001} argued that the expected SBPs are too uniform to reproduce
a significant microlensing deviation.  This conclusion appears to be
partially due to the fact that he found a best-fit impact parameter of
$b\sim 2$, whereas we find $b=0.44$--$0.65$ (and $b=0.65$ for our best fit
model).  All else equal, more uniform SBPs will result in microlensing
deviations that are shallower and broader than more limb-brightened SBPs.
Thus, to reproduce the observed deviation, a more uniform source must have
a smaller impact parameter, a trend we find empirically when fitting the
theoretical SBPs.  Thus, it is unclear why \citet{pana2001}, who argued that
the expected SBPs should be more uniform than that found by GLS, derived a value of $b$
that is twice as large ($b\sim 2$ versus $b\sim 1$).

We conclude that the microlensing of realistic SBPs can explain the
optical/infrared light curve of GRB 000301C. Fitting the data with a
non-parametric SBP reveals that the afterglow image must be significantly
limb-brightened. Specifically, $\gsim 60\%$ of the flux must originate from
the outer 25\% of the area of the image (with a significance of
$3\sigma$).  The recovered SBP is marginally consistent (at the $\sim
3\sigma$ level) with the profile expected for a uniform external density
profile ($k=0$) and frequencies $\nu_m<\nu<\nu_c$. It is consistent at the
$\sim 2\sigma$ level with the emission expected from frequencies $\nu >
\nu_c$ and both uniform ($k=0$) and wind-like ($k=2$) external media.

\section*{Acknowledgements}

We thank K. Stanek for providing an updated compilation of the data points.
This work was supported in part by NASA through a Hubble Fellowship grant
from the Space Telescope Science Institute, which is operated by the
Association of Universities for Research in Astronomy, Inc., under NASA
contract NAS5-26555 (for S.G.), by the Horowitz foundation and US-Israel
BSF grant BSF-9800225 (for J.G.) and the US-Israel BSF grant BSF-9800343,
NSF grant AST-9900877, and NASA grant NAG5-7039 (for A.L.).


\begin{thebibliography}{}

\bibitem[Berger et~al.(2000)]{berger2000}
Berger, E., et~al.\ 2000, ApJ, 545, 56

\bibitem[Blandford \& McKee(1976)]{bm1976} Blandford, R. D., \& McKee, C. F. 1976, Phys. Fluids, 19, 1130

\bibitem[Chevalier \& Li(2000)]{cl2000} Chevalier, R.A. \& Li, Z.Y. 2000, ApJ, 536, 195

\bibitem[Frail et~al.(2000)]{fwk2000}
Frail, D.A., Waxman, E., \& Kulkarni, S.R.\ 2000, ApJ, 537, 191

\bibitem[Frail et~al.(2001)]{frail2001}
Frail, D.A., et~al.\ 2001, Nature, submitted (astro-ph/0102282)

\bibitem[Freedman \& Waxman(2001)]{fandw1999} Freedman, D. L., \& Waxman, E. 2001, ApJ, 547, 922

\bibitem[Gaudi \& Loeb(2001)]{gandl2001} 
Gaudi, B.S., \& Loeb, A.\ 2001, ApJ, 558, 000 (astro-ph/0102003)

\bibitem[Garnavich, Loeb \& Stanek(2000)]{gls2000} 
Garnavich, P.M., Loeb, A., \& Stanek, K.Z.\ 2000, ApJ, 544, L11 (GLS)

\bibitem[Granot \& Loeb(2001)]{gl2001} Granot, J., \& Loeb, A. 2001,
ApJ, 551, L63

\bibitem[Granot et~al.(2001)]{granot2001}
Granot, J., et~al.\ 2001, to appear in the proceedings of the 2nd
workshop on 'Gamma-Ray Bursts in the Afterglow Era', Rome 2000
(astro-ph/0103038)

\bibitem[Granot, Piran \& Sari(1999a)]{gps1999} Granot, J., Piran, T.,
\& Sari, R.\ 1999a, ApJ, 513, 679

\bibitem[Granot, Piran \& Sari(1999b)]{gps1999b}
-------------------------------.\ 1999b, ApJ, 527, 236

\bibitem[Granot \& Sari(2001)]{gs2001}
Granot, J., \& Sari, R.\ 2001, in preparation

\bibitem[Harrison et al.(1999)]{harrison1999}
Harrison, F.A., et~al.\ 1999, ApJ, 523, L121

\bibitem[Ioka \& Nakamura(2001)]{iandn2001} 
Ioka, K., \& Nakamura, T.\ 2001, ApJL, in press (astro/ph-0102028)

\bibitem[Jensen et~al.(2001)]{jensen2001}
Jensen, B.L., et~al.\ 2001, A\&A, 370, 909

\bibitem[Koopmans \& Wambsganss(2001)]{kandw2001} Koopmans, L.V.E., \&
Wambsganss, J.\ 2001, MNRAS, submitted (astro-ph/0011029)

\bibitem[Kumar \& Panaitescu(2000)]{kandp2000}
Kumar, P., \& Panaitescu, A.\ 2000, ApJ, L541

\bibitem[Loeb \& Perna(1998)]{landp1998} Loeb, A., \& Perna, R.\ 1998,
ApJ, 495, 597

\bibitem[Mao \& Loeb(2001)]{mandl2001} Mao, S., \& Loeb, A.\ 2001, ApJ, 547, L97

\bibitem[Masetti et~al.(2000)]{masetti2000} Masetti, N., et~al.\ 2000, A\&A, 359, L23

\bibitem[Panaitescu(2001)]{pana2001} Panaitescu, A.\ 2001, ApJ, in press (astro-ph/0102401)

\bibitem[Panaitescu \& Kumar(2001)]{pandk2001} Panaitescu, A., \& Kumar, P.\ 2001, ApJ, in press (astro-ph/0010257)

\bibitem[Panaitescu \& Meszaros(1998)]{pm1998} Panaitescu, A., \& Meszaros, P.\ 1998, 493, L31

\bibitem[Piran(2000)]{piran2000}
Piran, T.\ 2000, Physics Reports, 333, 529

\bibitem[Rhoads \& Fruchter(2000)]{randf2000} 
Rhoads, J., \& Fruchter, A.S.\ 2000, ApJ, 546, 117

\bibitem[Sagar et~al.(2000)]{sagar2000} 
Sagar, R., Mohan, V., Pandey, S.B., Pandey, A.K., Stalin, C.S., \& Tirado, A.J.\ 2000, BASI, 28, 499

\bibitem[Sari (1998)]{sari1998} Sari, R. 1998, ApJ, 494, L49

\bibitem[Sari, Piran, \& Halpern(1999)]{sph1999} Sari, R., Piran, T., \& Halpern, T. 1999, ApJ, 519, L17

\bibitem[Sari, Piran, \& Narayan(1998)]{spn1998} Sari, R., Piran, T., \& Narayan, R. 1998, ApJ, 497, L17

\bibitem[Stanek et~al.(2000)]{stanek2000}
Stanek, K.Z., Garnavich, P.M., Barmby, P., \& Jha, S.\ 2000, BCN Circ.\ 766 
(http://gcn.gsfc.nasa.gov/gcn/gcn3/766.gcn3)

\bibitem[Van Paradijs et al.(2000)]{vp2000} 
Van Paradijs, J., Kouveliotou, C., \& Wijers, R.A.M.J. 2000, ARA\&A, 38, 379

\bibitem[Wijers \& Galama(1999)]{wg1999} Wijers, R. A. M. J., \& Galama, T. J. 1999, ApJ, 523, 177

\bibitem[Waxman(1997)]{wax1997c} Waxman, E.\ 1997, ApJ, 491, L19

\end{thebibliography}
\end{document}